\documentclass{aa}
\usepackage{graphicx}
\usepackage{txfonts}
\begin{document}
%

   \title{Maximum-likelihood detection of sources among Poissonian noise}

   \author{I. M. Stewart
          \inst{1}
          }

   \offprints{I. M. Stewart}

   \institute{Jodrell Bank Centre for Astrophysics, University of Manchester,
             Oxford Road, Manchester M13 9PL, United Kingdom\\
             \email{Ian.Stewart-2@manchester.ac.uk}
             }

   \date{Received January 0, 0000; accepted January 0, 0000}

   \abstract{
   A maximum likelihood (ML) technique for detecting compact sources in images of the x-ray sky is examined. Such images, in the relatively low exposure regime accessible to present x-ray observatories, exhibit Poissonian noise at background flux levels. A variety of source detection methods are compared via Monte Carlo, and the ML detection method is shown to compare favourably with the optimized-linear-filter (OLF) method when applied to a single image. Where detection proceeds in parallel on several images made in different energy bands, the ML method is shown to have some practical advantages which make it superior to the OLF method. Some criticisms of ML are discussed. Finally, a practical method of estimating the sensitivity of ML detection is presented, and is shown to be also applicable to sliding-box source detection.

   }

   \maketitle

\section{Introduction}

X-ray observatories from HEAO-2/Einstein onward have carried imaging cameras which have counted individual x-ray photons as they impinged upon a rectilinear grid of detector elements. The number of x-ray events detected per element within a given time interval will be a random integer which follows a Poisson probability distribution.

As the sensitivity of detectors and the effective areas of x-ray telescopes have increased, so has the x-ray cosmic background been increasingly resolved into compact sources. Eg a telescope such as XMM-Newton detects on average 70 serendipitous sources within the field of view of each pointed observation (Watson et al \cite{2xmm_paper}). A full characterisation of the x-ray background would shed light upon the evolution of the early universe (see Brandt and Hasinger \cite{brandt_2005} for a recent review of deep extragalactic x-ray surveys and their relation to cosmology). It is therefore of interest to find the best way to detect compact sources in these images - that is, the detection method which offers the best discrimination between real sources and random fluctuations of the background.

In a previous paper (Stewart \cite{stewart_2006}, hereinafter called paper I) I discussed the general class of linear detection methods, ie methods which form a weighted sum of the values of several adjacent image pixels, and optionally also over several different energy bands. A method for optimizing the weights was presented and shown to be substantially superior to the sliding-box method which, no doubt because of its simplicity, has been frequently used for compiling catalogs of serendiptitous sources (see eg DePonte \& Primini \cite{pros}; Dobrzycki et al \cite{celldetect}; also task documentation for the exsas task \emph{detect} and the SAS task \emph{eboxdetect}).

The present paper should be considered as an extension to paper I and also partly as an emendation. Several issues discussed in paper I either had not at that time been completely thought through, or were not described as clearly as they might have been. These issues are itemized as follows:

\begin{itemize}
  \item The difference between the distribution of samples of a random function, and the distribution of its peak values, was not realized.
  \item The concept of sensitivity was not adequately defined.
  \item The normalization of the signal did not allow direct comparison between single-band and multi-band detection sensitivities.
  \item Both paper I and the present paper are concerned not only with numerical values of the sensitivity of various methods of source detection, but also with practical, approximate methods of calculating these values `in the field'. One may prefer a lengthy, complicated method for calculating values for a paper; such is unlikely to be an equally satisfactory field method. In paper I there was insufficient grasp of this conceptual distinction.
  \item Sensitivities for the $N$-band \emph{eboxdetect} method were incorrectly calculated (see section \ref{sd4_cashbox} for an explanation).
\end{itemize}

The different signal normalization adopted in the present paper precludes direct comparison of the results herein with those of paper I; however after appropriate rescaling the two sets of results are consistent. No numerical errors in paper I were found, despite extensive reworking of the software for the present paper.

The primary aim of the present paper is to present a method of source detection based on the Cash statistic (Cash \cite{cash_1979}) and to show that this is at least as sensitive as the optimized linear filter described in paper I. For this purpose, sensitivities of a variety of detection methods are compared across a range of background values for both single-band and multi-band detection scenarios. The paper also aims to demonstrate other advantages of the Cash source detection, and to discuss and investigate some reported disadvantages.

Section \ref{sd2_srcdet} contains a mathematical description of the general source detection problem. In section \ref{sd2_3partalgo} a 3-part detection algorithm is described, in which a map of the detection probability at each image pixel is made as step one, peaks (maxima) in this map are located in step two, and a source profile is fitted to each peak in step three. The relation between the respective probability distributions of the map vs peak values is also discussed. Detection sensitivity is defined in section \ref{sd2_sensy}.

Cash-statistic source detection is described in section \ref{sd2_cash_def}. In section \ref{sd2_cash_mapvspeak} it is shown how the Cash statistic may be used at both stages 1 and 3 of the 3-part source detection algorithm. Finally, in sections \ref{sd2_cash_objections} and \ref{sd2_cash_verification}, objections to the Cash method are addressed.

Section \ref{sd3} contains the results of the sensitivity comparisons. In sections \ref{sd3_model} and \ref{sd3_technique}, details of the necessary calculations are discussed. 1-band results are presented in section \ref{sd3_1band_comp} and $N$-band results in section \ref{sd3_nband_comp}. The degradation of some $N$-band sensitivities when the true source spectrum does not match that assumed is discussed in section \ref{sd3_nband_nonmatching}.

An approximation useful for `in the field' estimation of Cash sensitivities is presented in section \ref{sd4_cash_approx}. Finally, in section \ref{sd4_cashbox}, a deficiency of the sliding-box method is discussed, and a method for remedying it is proposed.

\section{Theory}
\subsection{Source detection} \label{sd2_srcdet}

In the situation relevant to the present paper we have a model function $e(\vec{r})$ defined over the space $\vec{r}$ by
\begin{displaymath}
  e(\vec{r}) = B(\vec{r}) + \sum_j^M \alpha_j S_j(\vec{r}-\vec{r}_{0,j})
\end{displaymath}
where $B$ is called the background, each $\alpha$ is a scalar amplitude and $S$ is the signal shape. We conceive of a set of $N$ positions $\vec{r}_i$ for $i \in [1,N]$, each with an associated measurement of flux $c_i$. Each $c_i$ is a random integer with a Poisson distribution about a mean given by the value $e_i$ of the model at that position. From this comes the requirement that $e_i \ge 0 \ \forall i$. In mathematical shorthand
\begin{displaymath}
  \langle c_i \rangle = e_i = B_i + \sum_j^M \alpha_j S_j(\vec{r}_i-\vec{r}_{0,j}).
\end{displaymath}
In CCD x-ray detectors the $c_i$ are obtained by accumulation within arrays of voxels, ie volumes within the physical coordinates $\vec{r}$. In this case the functions $B$ and $S$ are integrated over the $i$th voxel dimensions to give the respective expectation values $B_i$, $S_{i,j}$ etc. For the purposes of the present paper it is mostly assumed that the coordinates $\vec{r}$ consist of just the spatial $(x,y)$ coordinates of the x-ray detector; in some sections the energy of the x-ray photons is added to the set. In this situation each voxel is bounded in the $(x,y)$ directions of course by the edges of the physical pixels of the detector. In the energy direction, voxels are bounded by the boundaries of any energy selection performed by the user.


The ultimate aim of source detection is to obtain the best estimates possible of $\alpha$ and $\vec{r}_0$ for the most numerous possible subset of the $M$ sources present. Satisfactory performance of this depends upon a couple of conditions. Firstly, the sources must not be confused - ie the density of `detectable' sources should be $\ll 1/\Delta\vec{r}_S$ where $\Delta\vec{r}_S$ is some characteristic size of $S$. For XMM-Newton this requires the detectable source density to be less than about 10$^5$ deg$^{-2}$. The deepest surveys to date reach x-ray source densities of only a tenth of this value (Brandt and Hasinger \cite{brandt_2005}). On the other hand, it was shown in paper I that perturbations of the background statistics set in at much shallower levels of sensitivity; but still a little beyond the deepest practical reach of XMM-Newton. There appears thus to be still some scope to improve source detection in the present observatories without coming up against the confusion limit. To some extent also one can disentangle confused sources during the fitting process, by comparing goodness-of-fit for fitting one versus two or more separate source profiles to a single detection.

If we may neglect confusion then we can detect sources one at a time. The model formula can thus be simplified to
\begin{equation} \label{equ_basic}
  e_i = B_i + \alpha S(\vec{r}_i-\vec{r}_0).
\end{equation}

The second condition necessary to source detection is that the form of $S$ is (i) known (at least approximately) and (ii) not too similar to $B$. In x-ray detectors used to date, a significant fraction of sources have angular sizes which are smaller than the resolution limit of the telescope. $S$ for such sources is then simply the PSF of the telescope, which can be estimated \emph{a priori}, and which is often much more compact, ie with more rapid spatial variation, than the background. However, about 8 percent of sources detected by XMM-Newton for example are resolved or extended (Watson et al \cite{2xmm_paper}). Detection of a resolved source is much more difficult (or at least, less efficient) since its $S$ is dominated by the angular distribution of source flux, which is not known \emph{a priori}. In addition, there is of course no upper limit to the size of x-ray sources, and in fact the distinction between source and background is largely a philosophical one. In practice one has to impose an essentially arbitrary size cutoff and consider only variations in x-ray flux which have spatial scales smaller than that cutoff to be sources, and take everything else to be background.

Modern CCD detectors can measure not only the positions of incident x-ray photons but also their energies. This opens an additional degree of freedom for $S$ and $B$ and thus the possibility of exploiting differences in their respective spectra to separate them with even greater efficiency. The average spectrum of the sources in the 1XMM catalog is quite different from the instrumental background spectrum of typical instruments (see Carter and Read \cite{carter_2007} and Kuntz and Snowden \cite{kuntz_2008} for recent data on XMM-Newton), and appears also to be different to the spectrum of the cosmic background (see eg Gilli et al \cite{gilli_2007}). However if it is true, as most authorities now suppose, that practically the whole of the cosmic x-ray background at energies higher than a few keV is comprised of point sources, one might expect the latter difference to diminish as fainter sources become detectable.

Parallel detection of sources in images in several energy bands of the same piece of sky is complicated by the fact that, although all (point) sources have the same PSF at a given detector position and x-ray energy, there is no corresponding common source spectrum. This means that we cannot `locate' the source $S$ in the energy dimension - we must be satisfied with locating it spatially on the detector.

Some of the detection algorithms described in the present paper avoid making any assumption about the source spectrum; others require the user to choose a spectrum \emph{a priori}. As will be seen from the results, there is usually a trade-off involved, in that methods which require the user to select a spectrum template may be better at detecting sources with spectra close to that template, but worse where the source spectrum is very different to the template; whereas a method which makes no assumptions will usually perform with about the same efficiency on all sources.

\subsection{Likelihood-map source detection} \label{sd2_3partalgo}

The present paper treats of a generic source-detection procedure which falls into three parts. The first part calculates a map giving an initial, possibly coarse estimate of the probability, for each pixel, that a source is centred on that pixel. In other words, it tests, for each $j$th map pixel, the following flux model:
\begin{equation} \label{equ_c}
  e_i = B_i + \alpha S(\vec{r}_i-\vec{r}_j).
\end{equation}
This task is accomplished by calculating some statistic $U$ from the count measurements $c_i$ taken from some subset of the $N$ detector pixels. The spatial distribution of this subset ought not to be much larger than the size of $S$ - otherwise one is including pixels which one knows contain no information about $S$. In other words, the pixels chosen to calculate the statistic at detector pixel $j$ are best located within a field of approximate size $\Delta\vec{r}_S$, centred about pixel $j$.

After calculating the statistic $U$ for all pixels, this first part of the detection procedure converts this into a measure of the probability that the value of $U$ at pixel $j$ would arise by chance fluctuation of the background alone.

The second part of the source-detection procedure consists of a peak detection algorithm. The output from this part is a list of peaks in the probability map, which are interpreted as source candidates.

The final task cycles over this list of candidates and attempts to fit the signal profile to the field of $c_i$ values surrounding each. To be of maximum use, the fitting procedure should be capable both of separating a single candidate into two or more confused sources, and merging multiple candidates which turn out to belong to a single source.

In the XMM-Newton data-product pipeline, both stages 1 and 2 are performed by the SAS task called \emph{eboxdetect}, whereas the final stage is performed by \emph{emldetect}. 

The final task may perform some additional calculations. It may calculate $\chi^2$ for the fit so as to aid in discriminating real cosmic sources from other phenomena, such as `hot' CCD pixels, which may give rise to excess counts (this is not at present done by \emph{emldetect}). As is shown in section \ref{sd2_cash}, use of the Cash statistic allows one to calculate a final value for the source-detection likelihood after the fit.

This separation of detection and fitting stages is a useful procedure only where the image pixels are not significantly larger than the characteristic width $\Delta\vec{r}_S$ of $S$. This criterion is in fact achievable with the imaging cameras of all three current observatories primarily dedicated to detecting x-rays, namely XMM-Newton, Suzaku and Chandra (Str\"uder et al \cite{strueder} and Turner et al \cite{turner} for XMM-Newton; Serlemitsos et al \cite{serlemitsos_2007} and Koyama et al \cite{koyama_2007} for Suzaku; and Weisskopf et al \cite{weisskopf_2002} for Chandra).

\subsubsection{Peak detection} \label{sd2_3partalgo_peak}

The peak-detection stage does not seem to have been the subject of much study. It is problematical for two reasons. Firstly, the frequency distribution of values at the maxima of a random function (call it the `peak' distribution) is not in general the same as the distribution of values over a set of randomly chosen positions (which we might call the `whole-map' distribution). In general one would expect the former distribution to be somewhat broader than the latter, since the probability that a random pixel is a peak is small for relatively low values but large for high values. Nakamoto et al (\cite{nakamoto_2000}) give a method to calculate the distribution of peaks for a random function of one dimension, but it is not clear if this theory can be extended to more than one dimension. However, as is shown in section \ref{sd2_cash_verification}, the Cash maximum likelihood statistic does provide a way to estimate the shape of the peak distribution, although not its normalization (which depends on the fraction of map pixels which are peaks). The normalization must be estimated by appropriate simulations.

The second problem is that the definition of a peak in an array of samples of a nominally continuous function is somewhat arbitrary. One obvious choice of definition is to label any $j$th pixel as a peak pixel if the value $c_j$ at that pixel is greater than the values at all other pixels within a patch of pixels surrounding the $j$th. But what size and shape to make the patch? It seems clear that these ought to match the size and shape of $S$ - too large a patch might miss additional, nearby sources; whereas too small runs into the opposite danger of listing more than one peak for a single source. Also, it is not clear how this choice might affect the frequency distribution of peak values. This danger was pointed out by Mattox et al (\cite{mattox_1996}) in respect of source detection in EGRET data.

\subsection{Sensitivity} \label{sd2_sensy}

The efficiency of a method of detecting sources is characterised by its sensitivity, which, broadly speaking, is the smallest amplitude $\alpha$ which will result in a source signal $\alpha S$ being detected among background of a given level. While this sentence conveys the general idea, a precise mathematical definition is necessary before a comparison can be done between methods of source detection. A somewhat stilted and obscure definition of sensitivity was presented in paper I. The present section is an effort to make this definition clearer and more rigorous.

\subsubsection{Definition} \label{sd2_sensy_definition}

Let us start by assuming that $B$ and $S$ are known \emph{a priori}. Formally speaking this is not true: one ought to make use only of their best estimates $\hat{B}$ and $\hat{S}$ in the formula for any source detection statistic. In practice however, because $B$ and $S$ can often be quite well estimated by methods separate from the source detection itself, it is convenient for the sake of simplicity to make the approximation $\hat{B} \sim B$ etc.

Each of the detection methods discussed in either paper I or in the present paper can be associated with a statistic $U$ which is to be evaluated at positions of interest: perhaps at the centre of each image pixel, or perhaps at a set of positions of candidate sources. $U$ at any position is evaluated, using a formula particular to the detection method, from the measured counts $c_i$. Clearly $U$ so calculated is a random variable which will occur with some well-defined probability distribution $p(U)$. We can also define an integrated (more strictly, a reverse-cumulative) probability $P$ such that
\begin{displaymath}
  P(>\!\!U) = \int_{U}^{\infty} dU \ p(U).
\end{displaymath}

In order to understand how sensitivity relates to $U$ it is helpful to imagine an ensemble of measurements of a source of a given amplitude $\alpha$. Or one could simulate this situation via a Monte Carlo experiment which, for each member of its ensemble, generates a set of $c_i$ values. The input model for this should consist of equation \ref{equ_c} and can make use of the \emph{a priori} known shapes of $B$ and $S$. For each member of the ensemble, $U$ is calculated from the $c_i$ values. The resulting ensemble of $U$ values will have a distribution which will in general be different for different \emph{a priori}-assumed values of $\alpha$. To reflect this, let us write both the probability density $p$ and the reverse-cumulative probability $P(>\!\!U)$ with an additional $\alpha$ argument as $p(U;\alpha)$ and $P(>\!\!U; \alpha)$.

We are interested first of all in the predicted $P$ in the case that $\alpha$ equals zero - ie, when there is no source in the detection field, only background. It is necessary to consider this scenario because the random nature of the data means that, at least some of the time, we will unavoidably be labelling as a source a concentration of counts which is merely a random fluctuation of the background.

The first step in source detection is to decide what fraction of these false positives we are prepared to put up with. This decision defines a cutoff probability $P_\mathrm{det}$ which in turn is associated (through the function $P(>\!\!U; 0)$) with a definite value of $U_\mathrm{det}$.

As an aside, note that in all cases in which the measured data are random integers, $p(U;\alpha)$ is a sum of delta functions, thus $P(>\!\!U; 0)$ is a descending step function. Inversion of the latter to obtain $U_\mathrm{det}$ can clearly only give an approximation in which $U_\mathrm{det}$ is set to the next highest value of $U$ for which $p$ is defined. In this situation one may expect the sensitivity of the method to descend in steps with decreasing background. This effect can be seen in the plots of the sensitivity of the simple sliding-box detection scheme in figures \ref{fig_a}, \ref{fig_e} and \ref{fig_i}.

\begin{figure}
   \centering
      \resizebox{\hsize}{!}{\includegraphics[angle=-90]{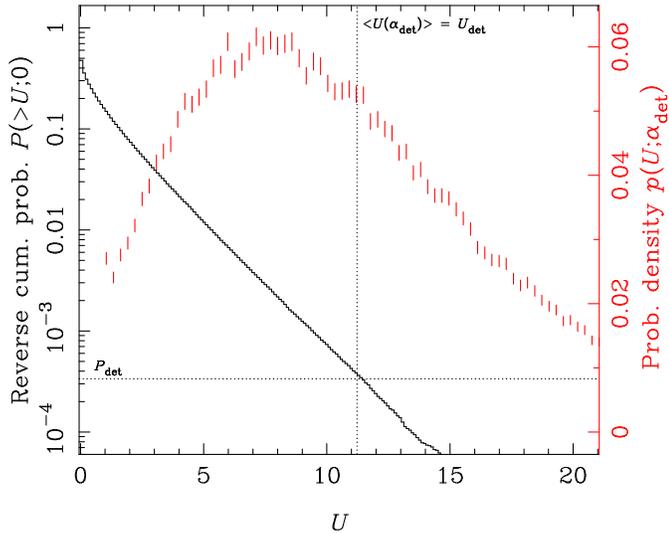}}
      \caption{The solid curve and the error bars (the latter shown in red in the electronic version) represent the results of separate Monte Carlo experiments, as described in the text. The solid curve shows the reverse-cumulative probability $P(>\!\!U;\alpha)$ in the case that the model consists entirely of background (ie, $\alpha = 0$). The horizontal dotted line shows the chosen detection probability $P_\mathrm{det}$. Its interception with the solid curve gives the cutoff value $U_\mathrm{det}$ of the detection statistic $U$. The error bars show the approximate course of the probability density function $p(U)$ for a case in which the model contains a non-zero source component. In the present case the model source amplitude $\alpha$ has been adjusted such that $\langle U(\alpha_\mathrm{det}) \rangle = U_\mathrm{det}$. This is the required condition for $\alpha$ to equal the sensitivity $\alpha_\mathrm{det}$.
              }
         \label{fig_p}
   \end{figure}

The sensitivity $\alpha_\mathrm{det}$ of a detection method is here defined to be that value of $\alpha$ which, for given $B$ and $S$, gives rise to values of $U$ such that $\langle U \rangle = U_\mathrm{det}$, where $U_\mathrm{det}$ is obtained from $P_\mathrm{det}$ as described above. See figure \ref{fig_p} for a diagrammatic example of this. The algorithm for calculating sensitivity is thus as follows:

\begin{enumerate}
  \item Decide a value for $P_\mathrm{det}$.
  \item Invert the relation $P = P(>\!\!U;0)$ (solid line in figure \ref{fig_p}) to obtain $U_\mathrm{det}$ from $P_\mathrm{det})$.
  \item An expression for the average value of $U$ as a function of alpha is inverted to give $\alpha_\mathrm{det} = <U>^{-1}(U_\mathrm{det})$.
\end{enumerate}

Note that other definitions are possible, for example one could define $\alpha_\mathrm{det}$ such that the median of $U(\alpha_\mathrm{det})$ instead of its mean was adjusted so as to equal $U_\mathrm{det}$. I have preferred the definition in terms of the mean simply because it is generally easier to calculate. However in the limit of high counts, all the varieties of $U$ here studied tend to a gaussian distribution, for which the mean and median are equal. Within the range of background values treated in the two papers, the median and mean definitions empirically are found to produce similar values.

The mean definition is equivalent to the `counts amplitude' concept employed in paper I; however the median definition is implicit in figure 12 of that paper.

The curves in figure \ref{fig_p} were constructed via Monte Carlo. At each iteration, simulated data values were generated for a small ($9 \times 9$) patch of pixels; the value of the appropriate detection statistic $U$ was then calculated for the patch. The end result of the Monte Carlo is an ensemble of $U$ values. The data values were Poisson-distributed integers whose mean values were given by a model comprising flat background of 1.0 counts/pixel plus a centred source profile. The solid curve and the error bars represent Monte Carlo-generated ensembles with $10^6$ and $10^5$ members respectively.

As mentioned in section \ref{sd2_3partalgo}, a multi-band detection method which does not require the user to make an \emph{a priori} choice of source spectrum is expected to perform better at `across the board' detection, because of the fact that cosmic x-ray sources don't adhere to a common spectral template. Note however that, regardless of the requirements of a (multi-band) detection method, calculation of its sensitivity \emph{always} requires the user to choose a spectral template. Eg in the Monte Carlo thought experiment described above, if $\alpha$ is non-zero, $S$ needs to be completely defined over the whole space $\vec{r}$, which in the multi-band case includes an energy dimension, in order to generate $c_i$ values and ultimately a value for $\langle U \rangle$.

\subsubsection{Map vs peak sensitivity} \label{sd2_sensy_mapvspeak}

In paper I it was implicitly assumed that if one made a frequency histogram $p_\mathrm{peak}(U)$ of $U$ values at an ensemble of locations of detected sources, that this would be related to the probability distribution $p_\mathrm{map}(U)$ of $U$ values at all image pixels by the following proportional relation:
\begin{displaymath}
  p_\mathrm{peak}(U) = \frac{A_\mathrm{image}}{A_\mathrm{beam}} p_\mathrm{map}(U)
\end{displaymath}
where $A_\mathrm{image}$ is the image area in pixels, and $A_\mathrm{beam}$ represents a constant known as the `beam area', taken to be approximately equal to the FWHM area in pixels of the PSF. If this were true, the cutoff probability $P_\mathrm{det}$ could be obtained by multiplying the maximum acceptable number of false detections per unit image area by the beam area. I now realize however that this assumption is not true: as described in section \ref{sd2_3partalgo_peak} and demonstrated in figure \ref{fig_o1}, the shapes of the two distributions are different. It follows then that one ought to use the probability distribution of $U$ values at peaks in the map, rather than at all map pixels, in calculating sensitivity.

This is a pity, because there are several attractive features of calculating `map' rather than `peak' sensitivities. Three of these have present relevance. The first and most obvious of these `attractions' is the retention of consistency with paper I.

Secondly, it is usually much easier and more straightforward to calculate `map' sensitivities. An adequate closed-form approximation for the `map' version of the cumulative probability function $P$ exists for at least some of the statistics considered herein, whereas the `peak' versions of $P$ must, for all but the Cash statistic, be estimated via a lengthy Monte Carlo. In addition, calculation of `peak' distributions naturally also requires the addition of a peak detector to the algorithm; the Monte Carlo must create a large image rather than a small patch of pixels; boundaries of same have to be avoided, and so forth.

The third thing which makes map rather than peak calculations attractive arises because the true number density of false detections (indeed of detections of any kind) depends not just on the formula for $U$ employed, but also on the nature of the peak-detection algorithm. Having to consider the effect of different algorithms within a `space' with two degrees of freedom rather than one is an unwelcome complication.

None of these nice features of the `map' calculation would justify its use, however, if it were not for the fact, as pointed out by Nakamoto et al (\cite{nakamoto_2000}), that the `map' and `peak' distributions of a random function become asymptotically similar towards high values of the function. This convergence is observed for example in figure \ref{fig_o1}. Since source detection by its nature deals with extreme values of the null-hypothesis probability function, it was felt to be acceptable to retain the `map' calculation for section \ref{sd3}, in which sensitivities of the various methods are compared. However when it comes to calculation of the absolute sensitivity of the Cash (or any other) statistic, of course the true `peak' version of $P$ must be employed.

\subsection{The Cash statistic} \label{sd2_cash}

\subsubsection{Definition} \label{sd2_cash_def}

Cash (\cite{cash_1979}) discussed a $\chi^2$-like statistic for Poisson data. Cash's statistic was constructed in two stages. The first stage comprises the log-likelihood expression
\begin{displaymath}
	L = \sum_{i=1}^{N} (c_i \ln[e_i] - e_i - \ln[c_i !])
\end{displaymath}
where $e_i$ is a model function for the expectation value $\langle c_i \rangle$. $L$ can be seen to consist of the logarithm of the product of the individual Poisson probability densities $p(c_i \mid e_i)$. If the model $e$ can be expressed in terms of one or more adjustable parameters, we can define the best fit values of these parameters as those which produce the largest value of $L$. Note that the factorial term is not a function of $e$ and thus can be neglected for fitting purposes.

The second stage of Cash's development was to form the difference
\begin{equation} \label{equ_a}
	U_\mathrm{Cash} = 2(\max[L_{q}] - L_\mathrm{null})
\end{equation}
where $L_\mathrm{null}$ is $L$ calculated by setting all parameters of the model $e$ to fixed, background values and $\max(L_{q})$ for integer $q$ is shorthand for the maximum result obtainable by varying a pre-selected number $q$ of parameters of $e$.

So long as the null hypothesis model can be expressed by a combination of legal values of the $q$ free parameters, $U_\mathrm{Cash} \ge 0$. According to Cash's theory, under this and other conditions discussed below, $U_\mathrm{Cash}$ in the asymptotic case is distributed approximately as $\chi^2_q$, ie $\chi^2$ for $q$ degrees of freedom.

Just as with $\chi^2$ for gaussian data, the Cash statistic can be used not only to fit parameter values of the model but also to obtain information about the probability distribution of those values. Cash for example uses it to derive confidence intervals for fitted parameters. The XMM-Newton SAS task \emph{emldetect} (online description at http://xmm.vilspa.esa.es/sas/current/doc/emldetect.ps.gz) employs the Cash statistic firstly to obtain best-fit values for the position, amplitude and optionally extent of sources in x-ray exposures, and secondly, at the conclusion of the fitting routine, to calculate a detection probability for each so-fitted source. Calculation of the detection probability is done by setting the source amplitude to zero in the expression for $L_\mathrm{null}$. The probability of obtaining the resulting value of $U_\mathrm{Cash}$ is then equivalent to the probability of obtaining the fitted values of the source parameters when there is nothing in the field but background.

Descriptions of the use of the Cash statistic for source detection can be found in Boese and Doebereiner (\cite{boese_2001}) and Mattox et al (\cite{mattox_1996}) to give just two examples. The expressions developed in the present paper are not based on either treatment but comparison reveals close similarities.

\subsubsection{Map vs peak Cash} \label{sd2_cash_mapvspeak}

In the present paper, the Cash statistic is evaluated under two different scenarios: either at the centre of each image pixel to make a map of $U_\mathrm{Cash}$ samples, or at the best-fit locations of detected sources. In the first case, the source is taken to be centred on the pixel in question (equation \ref{equ_c}): the amplitude is the only fitted (ie free) parameter. In the second case, the $x$ and $y$ coordinates of the source centre are fitted as well as the amplitude. It is as well to consider how the $U$ values returned by these respective situations are related.

In either case, practical considerations limit the number of image pixels included in the calculation of each value of $U_\mathrm{Cash}$ to a relatively small patch surrounding the pixel of interest. As can be seen from figure \ref{fig_d} however, the size of this patch has an asymptotically decreasing effect on the result of the calculation as this size becomes much larger than the characteristic size of the PSF. Consider then an ideal, continuous, two-dimensional function $U_\mathrm{ideal}$ obtained by calculating $U_\mathrm{Cash}$, with only the model amplitude free, at all points within the image boundary. Samples of this smooth function on the grid of points at the centres of the image pixels represent the asymptotic values of the actual, practical map-$U$ evaluation in the limit of large `patch' size. The values at peaks of $U_\mathrm{ideal}$ bear the same relationship to the actually calculated values of peak-$U$ at the end of the final fitting procedure. The addition of two more fitted parameters does not change the value of $U$ at these peaks: only the appropriate probability distribution.

These ideas can be expressed more compactly if we introduce a little notation. Let $U(A_\mathrm{map})_{i,j}$ be the `map' value of $U$ evaluated at the pixel centred on $(x_i,y_i)$, for a patch area $A_\mathrm{map}$; let $U(\hat{x}_0,\hat{y}_0,A_\mathrm{peak})$ be the value at the fitted source location $(\hat{x}_0,\hat{y}_0)$, for patch area $A_\mathrm{peak}$; and let $U_\mathrm{ideal}(x,y)$ be the ideal, smooth function. Then
\begin{displaymath}
	U_\mathrm{ideal}(x_i,y_i) = \lim_{A_\mathrm{map} \rightarrow \infty} U(A_\mathrm{map})_{i,j}
\end{displaymath}
and
\begin{displaymath}
	U_\mathrm{ideal}(\hat{x}_0,\hat{y}_0) = \lim_{A_\mathrm{peak} \rightarrow \infty} U(\hat{x}_0,\hat{y}_0,A_\mathrm{peak}).
\end{displaymath}

Imagine the 3-part detection scenario of section \ref{sd2_3partalgo} with the Cash prescription used both to calculate the map of $U$ values in step 1 and to fit the source parameters in step 3. As mentioned above, an advantage of the Cash statistic is that it can then be used to calculate a final value of $U$ (thus of the null-hypothesis probability) for each fitted source. If the image pixels are smaller than the characteristic size of the PSF (mentioned in section \ref{sd2_3partalgo} as a condition for the usefulness of the 3-part algorithm), and if we assume (as seems reasonable) that the maxima in $U_\mathrm{ideal}$ are asymptotically paraboloid, then we expect two things: firstly, that the positions output by step 2, ie positions of maxima in the array of `map' values produced by step 1, will be, most of the time, within 0.5 pixels of the final fitted positions; secondly, and because of this, that the final $U$ values will be approximately the same as, but slightly larger than, the $U$ values at the step 2 grid positions. These results are borne out by Monte Carlo experiments.

The second expectation, plus the fact that, as mentioned in section \ref{sd2_sensy}, the respective tails of the `map' and `peak' distributions of $U$ are expected to be similar, together mean that we may investigate the sensitivity of Cash statistic source detection, or at least compare it with other methods, via the (much more convenient) `map' process alone.

Since this paper shows that the Cash statistic is at least as sensitive as, and more flexible than, an optimized linear filter, it might be supposed that a scheme of source detection which employed Cash for stage 1 as well as stage 3 would be optimum in practice. However, although it has been found convenient to make use of this scheme as described above in the present paper, it can be shown that this is probably not the best scheme for practical source detection. In fact it isn't necessary to use the Cash statistic for step 1 as well as step 3; nor is this likely to be an optimum use of computer resources. An alternative method is well exemplified by the current XMM-Newton detection scheme, in which the detection statistic of step 1 is calculated using a sliding-box formula (see equation \ref{equ_1band_box_u}). The sliding-box statistic is much quicker to calculate than the Cash statistic, and its relatively poor sensitivity is offset by setting the pass threshhold low enough to pass any possible source through to step 3. Since step 3 does employ the Cash statistic, the full sensitivity of the latter is retained without a great penalty in computing time.

For purposes of the present paper, where the amplitude alone is fitted, this was done via a Newton-Rhapson method, as detailed in appendix \ref{app_cash}. Where more than one parameter was to be fitted, the Cash statistic was maximized via a Levenberg-Marquardt method (following the algorithm described by Press et al \cite{numerical_recipes}).

\subsubsection{Objections to ML measures of significance} \label{sd2_cash_objections}

Various authors (eg Freeman et al \cite{freeman_1999}, Protassov et al \cite{protassov_2002}, Bergmann and Riisager \cite{bergmann_2002}, Pilla et al \cite{pilla_2005} and Lyons \cite{lyons_2007}) have criticized the use of Maximum Likelihood statistics for the detection of sources or spectral lines in Poisson-noise data. There seem to be two main concerns: firstly that the parameter values which return the null hypothesis lie on the boundary of the space of permissable values; secondly that the data to which a line (or source) profile is fitted are also those used to estimate the background (null hypothesis) signal.

\begin{figure}
   \centering
      \resizebox{\hsize}{!}{\includegraphics[angle=-90]{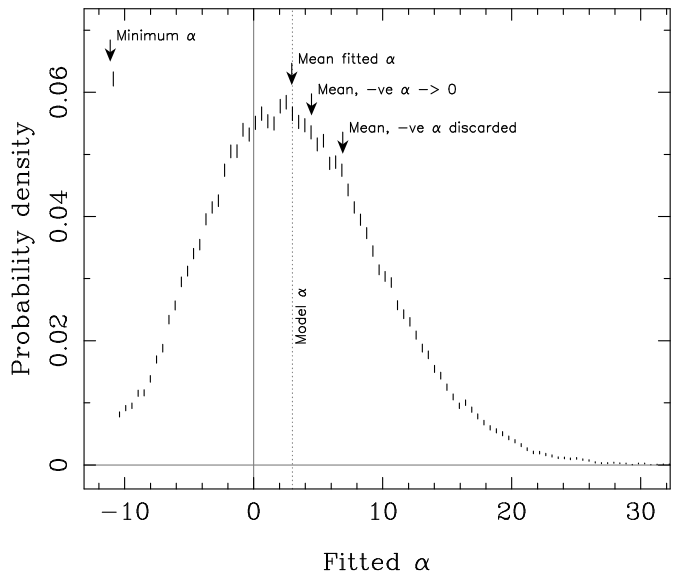}}
      \caption{The error bars are the result of a Monte Carlo experiment as described in the text. They show the frequency distribution of values of the source amplitude $\alpha$, fitted by use of the Cash statistic. All members of the Monte Carlo were generated using the same model amplitude (3.0 counts), indicated in this figure by the vertical dotted line. Negative values of fitted $\alpha$ were allowed; but the fit was not permitted to return a value less than the statistical limit (indicated with an arrow and the label `Minimum $\alpha$'), which is dictated by the necessity for the total model flux to remain $\ge 0$ for all pixels. The pile up of these bounded fits can be seen in the high numbers within the first bin. It can be clearly seen that it is necessary to retain the negative values of fitted $\alpha$ if the mean of the distribution is to be a good approximation to the model value of $\alpha$.
              }
         \label{fig_q}
   \end{figure}

The first objection appears to be a result of confusing physical with statistical constraints. Statistically speaking, for Poisson data, there is nothing wrong with a fit which returns a negative value for the amplitude, provided the net model flux $B + \alpha S$ remains non-negative. Even if one specifies that the physical model at issue does not allow negative amplitudes (although it is easy to conceive of situations which could give rise to a dip in flux at a particular position or energy), the fitted amplitude $\hat{\alpha}$ and the model amplitude $\alpha$ are not the same thing: the former is only an estimator for the latter, thus is a random variable with a certain frequency distribution. Where the model amplitude is small compared to the scatter in the fitted amplitude, it is necessary to allow negative values in order for the fitted amplitude to be an accurate estimator for the model. Figure \ref{fig_q} demonstrates this. (Figure \ref{fig_q} was constructed via Monte Carlo in a way similar to figure \ref{fig_p}. The ensemble comprised $10^5$ members.)

A negative value for $\hat{\alpha}$ which is small compared to its uncertainty $\sigma_\alpha$, where the physical model specifies that $\alpha \ge 0$, merely indicates that the measurement is consistent with zero. In this case one could use the sensitivity of the statistic as an upper limit. If $\hat{\alpha}$ is negative but large compared to $\sigma_\alpha$, the conclusion is rather that there is probably something wrong with the physical model (although, of course, such values will occur with a certain frequency by chance). In neither case are negative values of $\hat{\alpha}$ somehow incorrect. The \emph{only} requirement which the statistical theory imposes in cases where the noise is Poissonian is that the total flux (background plus source) may not be negative. Thus one can conclude, in contradiction to Protassov et al (\cite{protassov_2002}), that the null value of $\alpha$, ie zero, does not lie on the boundary of the permitted values of $\hat{\alpha}$.

Precisely what one does with negative $\hat{\alpha}$ values depends on one's aim. If that aim is to obtain the best estimate of $\alpha$, as in the above paragraphs, then negative $\hat{\alpha}$s should be retained. However if one desires to compile a list of candidate sources, then fits giving negative $\hat{\alpha}$ ought probably to be discarded from the list. In no case is it correct to keep the candidate but tamper with the $\hat{\alpha}$ value (by, for example, setting it to zero).

It should be pointed out that policy toward negative $\hat{\alpha}$ values should differ depending on whether the detection is over a single spectral band or over several in parallel. For single-band detection it is appropriate to discard candidates for which $\hat{\alpha}$ is negative - indeed, if a different algorithm is used for stage 1 of the detection process, as in the XMM-Newton pipeline, such candidates may never be submitted to Cash fitting in the first place.

The situation is slightly more complicated in the multi-band case. Because it is impossible to define a single source spectrum, it may be desirable to allow the flux in each band to be a free parameter - ie to fit the flux $\alpha_k$ in each $k$th band separately. Since one can easily conceive of a source which is bright in total but very faint in one of more spectral bands, it is entirely appropriate to retain negative values for $\hat{\alpha_k}$s in the final source list. As mentioned above, all that this implies is that the flux of that source in that band is undetectable.

Clearly, a criterion to decide whether a given multi-band, free-spectrum detection should be retained or not should involve some function of the fitted fluxes in all bands. It is just not quite clear what this function should be. Simplest may be to simply add the fluxes and retain only those sources for which the result is positive.

The validity of the second objection to maximum likelihood fitting, namely that the background and signal amplitude are obtained from the same set of data, depends on the circumstances of the detection. However, in XMM-Newton practice at least, the fraction of image area occupied by identifiable sources is usually small. In this case it seems a reasonable procedure to make an estimate of background from those regions of the image which one has concluded probably don't contain any detectable sources. The XMM-Newton pipeline performs a simple 2-step iterative procedure to calculate a background estimate by this method. Since one is in this case not using the same data for both background estimation and source fitting, the second objection does not apply.

Protassov et al propose the replacement of the ML statistic in these problems by a Bayesian technique. While admittedly more rigorous, their technique would require at least a fresh Monte Carlo to be performed for each source. Protassov et al in their spectral line examples deal in significances on the order of a few percent. To calibrate the distribution function of their detection statistic, a Monte Carlo of a few hundred elements suffices. Source detection however typically demands a much more stringent significance level: $3 \times 10^{-4}$ for 1XMM (Watson et al \cite{1xmm_paper}) for example. Monte Carlo ensembles with on order $10^5$ or more members are required to calibrate distributions down to this level. Such a technique is not practical for the compilation of a source catalog with upward of 50 000 members.

\subsubsection{Empirical verification of ML source detection} \label{sd2_cash_verification}

\begin{figure}
   \centering
      \resizebox{\hsize}{!}{\includegraphics[angle=-90]{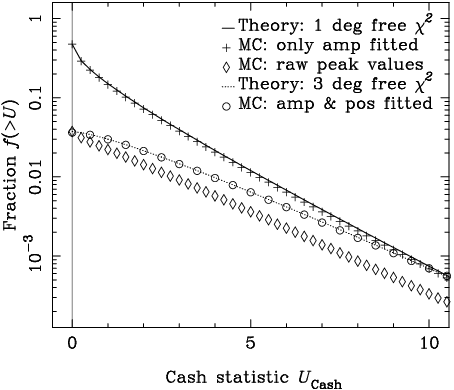}}
      \caption{Reverse-cumulative frequency histograms from Monte Carlo ensembles of the Cash statistic $U_\mathrm{Cash}$. Continuous curves plot the theoretical distributions, which are just the $\chi^2_n$ distributions for the respective $n$ degrees of freedom, suitably normalized to the respective fraction of pixels involved. Crosses, circles and diamonds represent Monte Carlo data. The solid curve (theory) and crosses (Monte Carlo) show results where only the source amplitude is fitted (= 1 degree of freedom), for a source centred at each image pixel. The diamonds give the distribution of $U$ values for that small subset of these `map' pixels which were identified as peaks. Finally, the dotted curve (theory) and the circles (Monte Carlo) show the distribution of values obtained by fitting $x$ and $y$ coordinates as well as amplitude (= 3 degrees of freedom) for each of the `map' peaks.
              }
         \label{fig_o1}
   \end{figure}

An empirical look at some Monte Carlo data shows indeed that the raw distribution of values of the Cash statistic does not appear to conform very closely to the ideal $\chi^2$ curve. This is so whether one sets the value of $U_\mathrm{Cash}$ to zero whenever the fitted amplitude is negative (as in the examples of Protassov et al) or whether one simply retains such values unaltered. However, discarding negative-amplitude fits from the pool of source candidates, as described in the preceding section, brings the distribution of the remainder very close to the $\chi^2$ theoretical curve. Figure \ref{fig_o1} shows the results of such a Monte Carlo.

\begin{figure}
   \centering
      \resizebox{\hsize}{!}{\includegraphics[angle=-90]{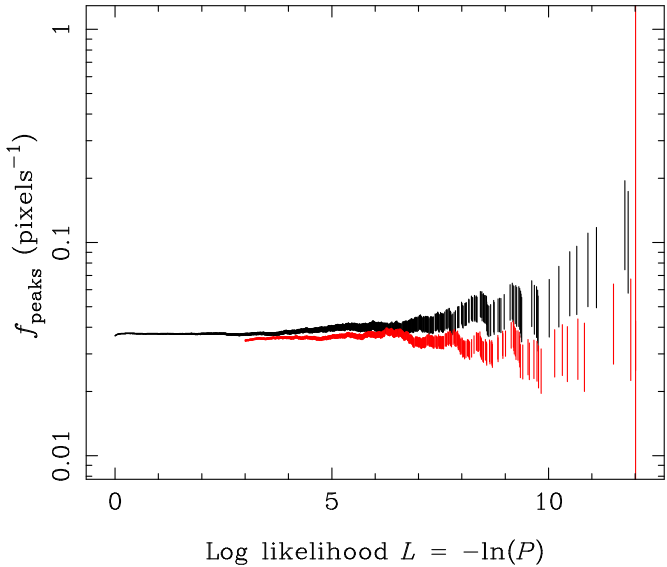}}
      \caption{A plot which allows the fraction of peak pixels $f_\mathrm{peaks}$ to be estimated. This value is just the height to which the curve asymptotes at small values of $L$. As described in the text, a flat slope for such a curve indicates that the shape (but not necessarily the normalization) of the empirical probability distribution agrees with that predicted by theory. The black bars show the results of the bespoke Cash fitting software; the half-tone bars (red in the electronic version) show the results of the SAS detection procedure. Both plots are truncated some way short of zero because only candidate sources having a `map' probability above a given cutoff were submitted to the fitting procedure. A number of computational reasons led to the choice of a much larger cutoff for the SAS fitting task (\emph{emldetect}).
              }
         \label{fig_m}
   \end{figure}

For figures \ref{fig_o1} and \ref{fig_m}, it was not possible to deal only with a small patch of simulated-data pixels, such as for figures \ref{fig_p} and \ref{fig_q}, because in the present case it was necessary to perform all three steps of the generalized source-detection algorithm (section \ref{sd2_3partalgo}). For this reason a much larger, circular field was used, with a radius (204 pixels) similar to that ($\sim 210$ pixels) of the field of view of the XMM-Newton EPIC cameras, when expressed in the standard 4-arcsec pixels of the data products. No sources at all were used in the input model for the Monte Carlos of figures \ref{fig_o1} and \ref{fig_m}: all detections depicted in these figures are `false positives'.

The penalty for improving the shape of the distribution by discarding negative fits is that one needs to estimate the expected fraction of such fits in order to normalize the distribution. When considering the statistics of `map'-stage Cash (ie fitting of amplitude alone at each pixel of an image), this fraction asymptotes to 0.5 in the limit of high background (see eg the zero intercept of the solid curve in figure \ref{fig_o1}). This value has been found to be an acceptable approximation throughout the range of background values used in the present paper.

Where the source position as well as amplitude is fitted using the Cash statistic, this simple approximation for the fraction of image pixels which yield candidates no longer applies. In this case one also needs to know the fraction of pixels which are peaks. There appears to be no way to estimate this apart from a Monte Carlo. One can derive the number \emph{a posteriori} via a plot such as figure \ref{fig_m}.

Figure \ref{fig_m} was created as follows. An ensemble of (false) source detections was created via a Monte Carlo as described above in the description of figure \ref{fig_o1}. The detection software makes an estimate of the probability $P$ that a source of that amplitude or greater will occur by chance from background. If that estimate is proportional to the true probability, then the proportionality may be expressed as follows:
\begin{displaymath}
	N(>\!\!P) \sim M_\mathrm{fields} A_\mathrm{field} f_\mathrm{peaks} P
\end{displaymath}
where $N(>P)$ is the number of sources in the ensemble having $P$ or greater, $M_\mathrm{fields}$ is the number of fields, $A_\mathrm{field}$ is the area of the field in pixels, and $f_\mathrm{peaks}$ is the fraction of image pixels which are identified as peaks (ie, positive-valued local extrema). (Note that the diamonds, circles and dotted line on figure \ref{fig_o1} should all trend to $f_\mathrm{peaks}$ as $U$ approaches zero.) Taking logs and manipulating gives
\begin{displaymath}
	\log_{10} \left( \frac{N}{M_\mathrm{fields} \ A_\mathrm{field}} \right) + \frac{L}{\ln(10)} = \log_{10}(f_\mathrm{peaks})
\end{displaymath}
where the replacement $L = -\ln(P)$ has been made since this is the value written by \emph{emldetect} to the column named DET\_ML in the output source list.

For each point on figure \ref{fig_m}, $N$ is calculated as the number of sources having that value of $L$ or higher. The error bars are just $\sqrt{N}$. Thus $N$ increments as one proceeds from the highest $L$ value to the lowest.

Results for two source-detection procedures are plotted on figure \ref{fig_m}. The black bars show results from software written for the present paper which performed all three stages of the source detection (section \ref{sd2_3partalgo}). The half-tone bars (red in the electronic version) result from use of the XMM-Newton SAS tasks \emph{eboxdetect} followed by \emph{emldetect}. Actual XMM-Newton product file headers were used in the simulated image, exposure map and background map files in order to facilitate acceptance of these by the SAS source detection tasks.

The asymptotic flatness of the figure \ref{fig_m} plots again supports the close correspondence between the actual distribution of values of the Cash statistic and the $\chi^2$ distribution predicted by the Cash theory.

Finally, although it is beyond the scope of the present paper to perform detailed tests on \emph{emldetect}, one can at least say from figure \ref{fig_m} that the value of DET\_ML is being calculated correctly. This seems not to be consistent with the results shown in figure 4 of Brunner et al \cite{brunner_2008}, which were also calculated from simulated data. However, their simulation was much more sophisticated: every field contained very many faint confusing sources, and a statistical procedure had to be used to identify `false positives'. This may explain the discrepancy with present results.

\section{Sensitivity comparison} \label{sd3}
\subsection{The model} \label{sd3_model}

The flux model used in the simulation was based on equation \ref{equ_c}. The physical coordinates here of interest are the two spatial coordinates $x$ and $y$, representing position on the detector (or, equivalently, position on a projection plane tangent to the celestial sphere), plus the energy $E$ of the x-ray photons. Samples along both $x$ and $y$ axes are taken to be equally spaced, located indeed on an orthonormal array, centred on the source centre $(x_0,y_0)$; but samples may be unevenly spaced in energy. It is convenient to modify equation \ref{equ_c} to make use of three indices, one per coordinate, viz:
\begin{displaymath}
	e_{i,j,k} = B_{i,j,k} + \alpha S_{i,j,k}.
\end{displaymath}
Both $i$ and $j$ are restricted to the range $-M$ to $M$ for some small integer $M$. The symbol $N$ is henceforth used to denote the number of energy bands.

The signal samples are related to the PSF $S(x,y,E)$ as follows:
\begin{displaymath}
	S_{i,j,k} = \int_{x_i-\Delta x/2}^{x_i+\Delta x/2} \int_{y_j-\Delta y/2}^{y_j+\Delta y/2} dx \ dy \ S(x,y,E_k)
\end{displaymath}
where $\Delta x$ and $\Delta y$ are the respective sizes of the pixels in the $x$ and $y$ directions and $E_k = (E_{k,\mathrm{lo}} + E_{k,\mathrm{hi}})/2$, ie the average of the bounding energies of band $k$. $S$ is normalized such that
\begin{equation} \label{equ_g}
	\frac{1}{\Delta x \Delta y} \ \int_{-\infty}^{\infty} \int_{-\infty}^{\infty} dx \ dy \sum_{k=1}^N S(x,y,E_k) = 1.
\end{equation}
(Note that this is not the same normalization condition as used for paper I.) Hence, $\alpha$ has units of x-ray counts and may be interpreted directly as the expected total number of counts, summed over all $N$ bands and over the entire detector plane, which will be generated by the source $S$ during a given exposure duration.

As for paper I, both $\Delta x$ and $\Delta y$ are set to the same value of 4 arcsec, which is the same as the pixel size of the x-ray images produced as standard XMM-Newton products, and as used for source detection in both the 1XMM and 2XMM catalogs (Watson et al \cite{1xmm_paper} and \cite{2xmm_paper}).

The PSF used, for both single-band and multi-band calculations, was that corresponding to the pn camera of XMM-Newton on the optical axis, and at a photon energy of 1.25 keV, which is the centre of band 2 of the 1XMM catalog x-ray data products (Watson et al \cite{1xmm_paper}). In paper I the PSF was allowed to vary in shape with energy band, as it is observed to do in XMM-Newton. This is discarded in the present paper in order to avoid confusion between the small effect due to the change in shape of $S$ with energy and the large effect due to its changing magnitude.

The expected background flux $B_{i,j,k}$ in counts per pixel is taken throughout to be spatially invariant, but in general different within each energy band. Source and background spectra used, plus the associated energy band definitions, are taken from table 1 of paper I.

\subsection{The sensitivity calculation} \label{sd3_technique}

As noted in section \ref{sd2_sensy}, the first step in source detection, thus also in calculating the sensitivity of that detection, is to decide the acceptable probability of a false detection. Here, as in paper I, a value of $\exp(-8.0)$ is used for this probability cutoff $P_\mathrm{det}$. This was the value used in the making of the 1XMM catalog (Watson et al \cite{1xmm_paper}).

The next step is to invert the appropriate expression for $P(>\!\!U;0)$ to obtain $U_\mathrm{det}$. In a minority of cases, an accurate closed-form expression for $P$ is known: in these cases, $P$ was inverted by use of Ridders' method (Ridders \cite{ridders}) as described in Press et al (\cite{numerical_recipes}). Ridders' method was chosen rather than a Newton-type method because of a paucity of expressions for $\partial P / \partial U$.

In the majority of cases, no closed-form expression or approximation for $P$ is known which is accurate enough for present purposes. For each detection method without an accurate formula for $P$, a Monte Carlo technique was used to estimate $U_\mathrm{det}$. An ensemble of $\eta$ sets of $c_{i,j,k}$ were generated from a model comprising purely the background $B_k$. $U$ was calculated for each set of counts values. The final ensemble of $U_m$ values, for $m \in [1,\eta]$, was sorted into increasing order. $U_\mathrm{det}$ is then approximately that value of $U_m$ for which $(\eta-m)/\eta \sim P_\mathrm{det}$. Clearly $\eta$ must be chosen such that $m \gg 1$.

Having obtained $U_\mathrm{det}$, the final step is to invert an expression for $\langle U \rangle$ to derive $\alpha_\mathrm{det}$. An accurate and invertible expression for $\langle U \rangle$ could be found for the majority of the detection methods. Where not, a Monte Carlo technique had to be used for this step of the calculation as well. In this onerous and somewhat problematic procedure, a Ridders iteration was employed to converge on a value of $\alpha_\mathrm{det}$. For each trial value in the iteration sequence, an ensemble of sets of $c_{i,j,k}$ was obtained. Each set yielded its value of $U$. The mean of the ensemble of $U$ values was tested against $U_\mathrm{det}$ and the trial $\alpha$ adjusted accordingly until convergence was judged to have been reached.

\subsection{1-band results} \label{sd3_1band_comp}

Three single-band detection methods are compared here, viz:

\begin{enumerate}
  \item Sliding-box: $U$ is obtained simply by summing the $c_i$ values within the box.
  \item Optimized linear filter: $U$ is obtained via a weighted sum of the $c_i$, the weights being chosen to optimize the sensitivity.
  \item Cash-statistic detection: $U$ given by the Cash statistic where amplitude only is fitted.
\end{enumerate}

\noindent
Formulas for $P(>\!\!U;0)$ and $\langle U \rangle$ are given in appendix \ref{app_PU_1band}.

\begin{figure}
   \centering
      \resizebox{\hsize}{!}{\includegraphics[angle=-90]{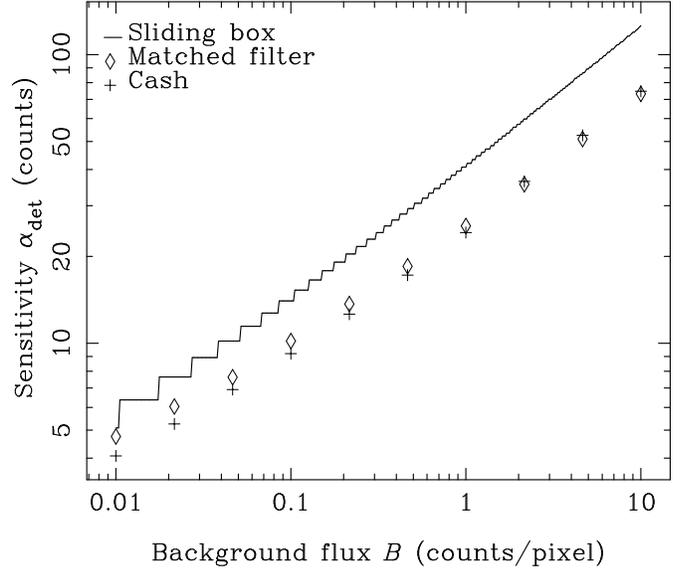}}
      \caption{Minimum detectable source counts $\alpha_\mathrm{det}$, plotted as a function of background flux $B$; single-band case. The solid line gives the results of the sliding-box technique; the diamonds show the results of detection by matched linear filter; the crosses show the results of the Cash method.
              }
         \label{fig_a}
   \end{figure}

Sensitivities of the three methods are compared for a range of values of background flux $B$ in figure \ref{fig_a}. All values were calculated for a square array of pixels with $M=4$.

The reason the sliding-box curve shows stepwise discontinuities is because (as described in section \ref{sd2_sensy}) $U$ for any method of detecting signals in Poissonian data must be a discrete function, not defined for all values on the real line, due to the discrete nature of the counts value in any pixel. All the $P$ functions, and any expression derived from them, thus descend in step fashion. This is most obvious for the sliding-box method because $U$ for this method is highly degenerate - ie there are many arrangements of counts $c_i$ which will produce the same $U$. The steps are still present for the other methods but are usually too finely spaced to be noticeable.

As one might expect, both the matched-filter and Cash methods are seen to be more sensitive (by a factor of at least 20\% at any value of background) than the sliding-box method. This is because they both make use of more information, namely the shape of $S$, to better discriminate between $S$ and $B$. What is not so expected is that the Cash method appears to be marginally better than the matched-filter method. At first it seemed possible that this was an artifact, a result of insufficient convergence of the Powell's-method routine which calculates the weights for the matched-filter method; however, tests using much stricter convergence criteria failed to reveal any significant dependence. The matched filter detection is by definition the optimum \emph{linear} filter; however the Cash detection algorithm is based however on a \emph{nonlinear} expression, which is optimized not just for the general conditions but for the particular arrangement of counts within each $(2M+1)$ by $(2M+1)$ detection field. One must assume that this accounts for its better performance.

Figure \ref{fig_a} is comparable to figure 6 of paper I, with the addition of the points for the Cash results. Some  differences can however be observed between the sensitivity values displayed in figure \ref{fig_a} and those given in the earlier figure. There are three reasons for this. Firstly, the normalization expression for $S$ is different: in paper I, the normalization integral for $S$ extended only as far as the boundary of the square $(2M+1)$ by $(2M+1)$ patch of pixels in which counts were considered; in the present paper the integral extends over the whole plane.

Secondly, a smooth approximation to $P(>\!\!U_\mathrm{box};0)$ was used in paper I, rather than the true step-function form used here.

\begin{figure}
   \centering
      \resizebox{\hsize}{!}{\includegraphics[angle=-90]{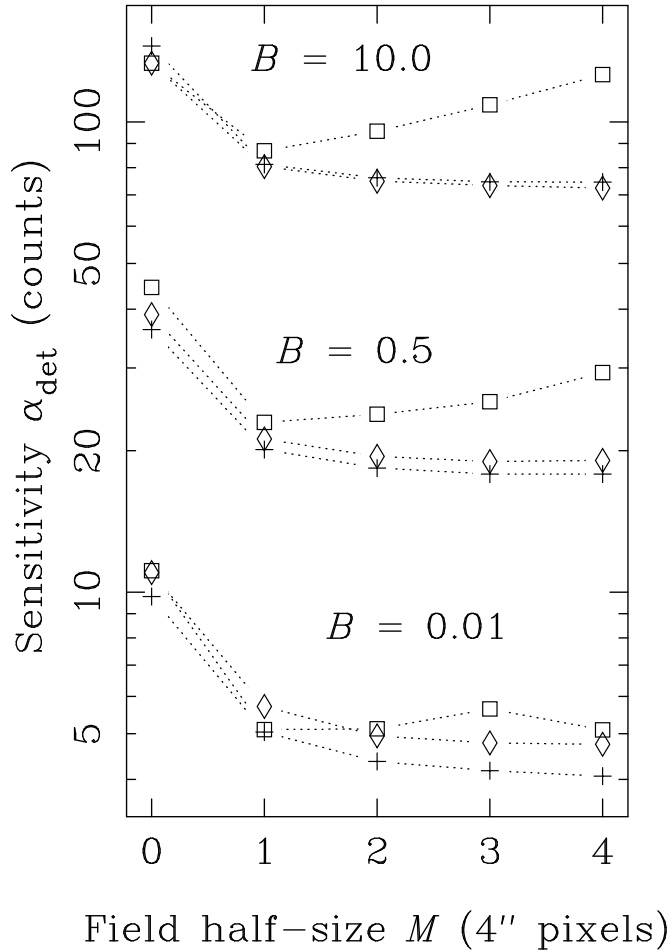}}
      \caption{The sensitivity of different detection methods as the size of the `detection field' is varied. The field side length is $2M+1$ pixels. Results are presented for three values of expected background flux $B$, given in counts per pixel. Those results belonging to the same method and the same value of $B$ have been linked with dotted lines to improve the legibility. The squares represent the sliding box convolver; diamonds, the matched filter method; and crosses, the Cash method.
              }
         \label{fig_d}
   \end{figure}

Thirdly, whereas in the present paper the same value of $M=4$ has been used for all methods, in paper I the values for the sliding-box method were calculated using $M=2$. Figure 4 in paper I was provided to allow the reader to estimate the effect of this choice. This figure is repeated here as figure \ref{fig_d}, with the Cash results included.

The points to note from figure \ref{fig_d} are twofold. The first is that, for both the matched-filter and Cash methods, the minimum detectable source count always decreases as the size of the field is increased, whereas the sliding box method reaches a minimum, with a slight dependence on the background level. The reasons for this were explored in paper I but, to reprise briefly: when confronted with a large field, most of which must be almost pure background, the sliding box method adds in all the background-dominated counts with equal weight, which therefore increases the total noise; whereas the matched-filter method assigns smaller weights to pixels as the fraction of $S$ to $B$ decreases. The same is true of the Cash method, for which the fit is dominated by pixels having a high $S/B$ ratio.

The second point to observe in figure \ref{fig_d} is that all methods become of roughly similar worth as $M$ becomes small. For $M=0$, corresponding to a detection field consisting of a single pixel, one would expect them to become identical, since any knowledge of the shape of $S$ is useless in this case. The reason for the persistence of some differences at $M=0$, particularly at low background, is unknown. However, the differences occur within a regime for which the expected total number of counts is less than 1. One might expect most measures to be a little `noisy' under such circumstances.

\subsection{$N$-band results} \label{sd3_nband_comp}

Five multi-band detection methods are compared here, viz:

\begin{enumerate}
  \item Sliding-box: $U$ is obtained simply by summing the $c_i$ values within the box, over all bands.
  \item \emph{eboxdetect}: this is also a sliding box method, but differs from the previous in that calculation of $U$ is two-stage. The first stage calculates $U_k$ for each $k$th band by summing values as usual. $U$ is then obtained from the $U_k$ via a formula (see appendix \ref{app_PU_nband_box_n}) for which the theoretical underpinning is unclear.
  \item Optimized linear filter: $U$ is obtained via a weighted sum of the $c_i$ over all pixels and bands, the weights being chosen to optimize the sensitivity. The complete form of $S$, thus of the source spectrum, must be assumed.
  \item Cash detection, fixed spectrum: $U$ is given by the Cash statistic where amplitude is fitted for all bands in parallel. A source spectrum must be assumed. There is thus only 1 degree of freedom in the fit.
  \item Cash detection, free spectrum: same as the preceding, except that the amplitude is fitted for each band separately, giving $N$ degrees of freedom, for $N$ bands.
\end{enumerate}

\noindent
Formulas for $P(>\!\!U;0)$ and $\langle U \rangle$ are given in appendix \ref{app_PU_nband}.

\begin{figure}
   \centering
      \resizebox{\hsize}{!}{\includegraphics[angle=-90]{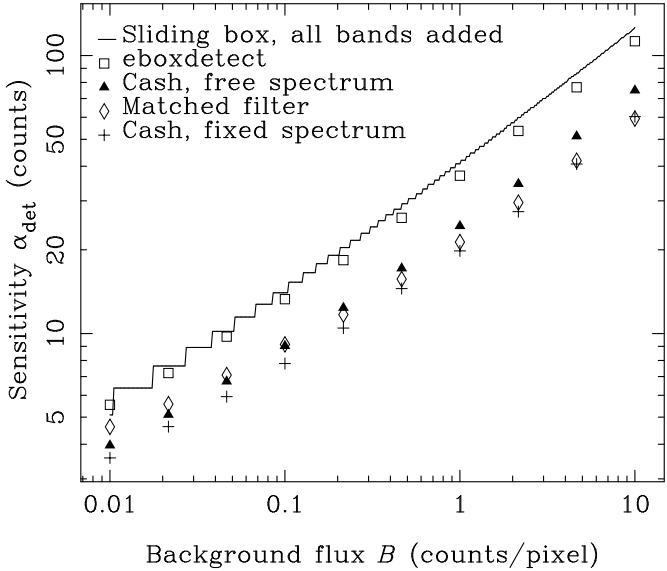}}
      \caption{Minimum detectable source counts $\alpha_\mathrm{det}$, plotted as a function of background flux $B$; $N$-band case. The solid line gives the results of the technique in which all bands are summed before sliding-box is applied (method 1 this section). Squares represent the results of the \emph{eboxdetect} sliding-box technique (method 2). Diamonds show the matched-filter results (method 3). The Cash method in which a source spectrum is assumed (method 4) is represented by crosses; the alternate, assumption-free Cash method (method 5) is represented by filled triangles.
              }
         \label{fig_e}
   \end{figure}

Sensitivities of the five $N$-band methods are compared for a range of values of background flux $B$ in figure \ref{fig_e}. All values were calculated for a square array of pixels with $M=4$.

The background flux $B$ in all figures relating to multi-band detection was calculated by summing the background $B_k$ in each band. This provision, together with the normalization of $S$ described in equation \ref{equ_g}, allows one to compare directly single- to multi-band detection. This was not the case in paper I.

The same changes and improvements as described in section \ref{sd3_1band_comp} hinder direct comparison of figure \ref{fig_e} of the present paper with figure 9 of paper I. However if one increases all values in figure 9 paper I by about 40\% to compensate for the different normalization of $S$ in that paper, the respective curves (diamonds in both papers indicate the matched-filter results, whereas the Monte Carlo-corrected \emph{eboxdetect} results are indicated by crosses in paper I and squares in the present paper) are seen to match reasonably well.

The normalization condition for $S$ chosen in the present paper does however facilitate comparison between the 1-band and $N$-band situations. The summed, sliding-box detection method is seen, as expected, to provide identical results here as in the single-band case. Some of the other $N$-band methods are conceptually similar to 1-band counterparts, and have been given the same graph symbols to highlight this fact; however, only for the sliding-box method are the $N$-band and 1-band sensitivity values expected to be identical.

The first thing to note is that the \emph{eboxdetect} version of sliding-box, in which the likelihoods for each band are summed to give $U$, performs slightly better than the simplistic summed-band version. Why? The answer is that the simple sum actually implies an assumption about the source spectrum, namely that it is flat - or, more precisely, that the source spectrum is such that the optimum weights per band are identical. This method might therefore be expected to perform better than the \emph{eboxdetect} method on sources which obey this criterion, and worse (as in the present case) when they do not. However there seemed little point to the present author in testing this speculation because the summed-likelihood sliding box method is in any case far from optimum.

Again the equivalent matched-filter and Cash methods, namely methods 3 and 4 above, represented respectively by diamonds and crosses, perform about the same. The improved perfomance of the Cash method at low values of background is no doubt due to the same reasons as in the single-band case.

As one might expect, the version of the Cash statistic which does not require an \emph{a priori} decision about the source spectrum performs less well than methods which do, at least in the present case in which the true source spectrum matches the template. The contrary case is examined in the next subsection.

\subsection{Effect of non-matching spectra on sensitivity} \label{sd3_nband_nonmatching}

\begin{figure}
   \centering
      \resizebox{\hsize}{!}{\includegraphics[angle=-90]{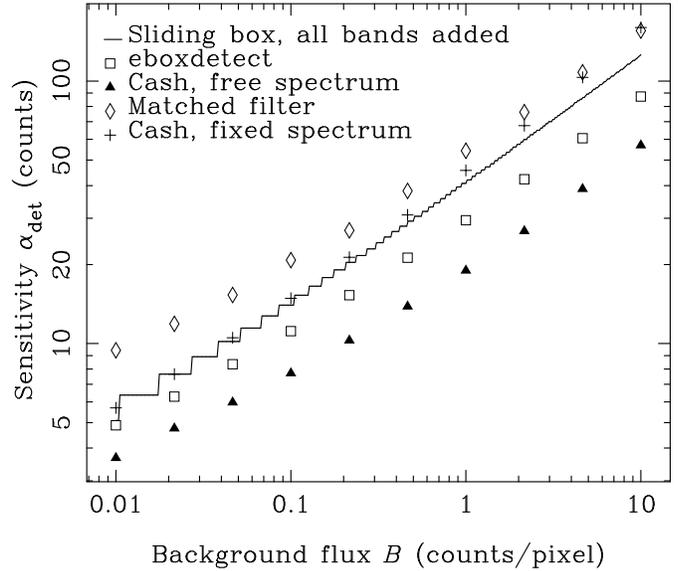}}
      \caption{Same as figure \ref{fig_e} except a hard source spectrum has been used to generate the data.
              }
         \label{fig_i}
   \end{figure}

Figure \ref{fig_i} is similar to figure \ref{fig_e}, except that there is now a mismatch between the spectrum used to generate the counts (and to calculate the sensitivity) and the spectrum assumed in the detection process. Such a mismatch is of course only meaningful for those methods which require a spectrum to be assumed. The hard source profile in table 1 of paper I was used to generate the Monte Carlo data, whereas the more representative source spectrum, the same used for both purposes in figure \ref{fig_e}, has been used in the appropriate detection schemes.

The summed-band sliding-box method again provides the unvarying benchmark. Among the remaining methods, two trends are evident: those methods (respectively \emph{eboxdetect} and free-spectrum Cash) which do not assume a source spectrum perform about the same as previously; whereas those which do (fixed-spectrum Cash and matched-filter) perform now significantly worse. This is as one would expect.

It is interesting that the fixed-spectrum Cash and matched-filter methods now diverge quite markedly at low values of background. The reason for this behaviour is not known.

\section{Some $\langle U \rangle$ and $P$ difficulties} \label{sd4}
\subsection{An approximate expression for $\langle U_\mathrm{Cash} \rangle$} \label{sd4_cash_approx}

\begin{figure}
   \centering
      \resizebox{\hsize}{!}{\includegraphics[angle=-90]{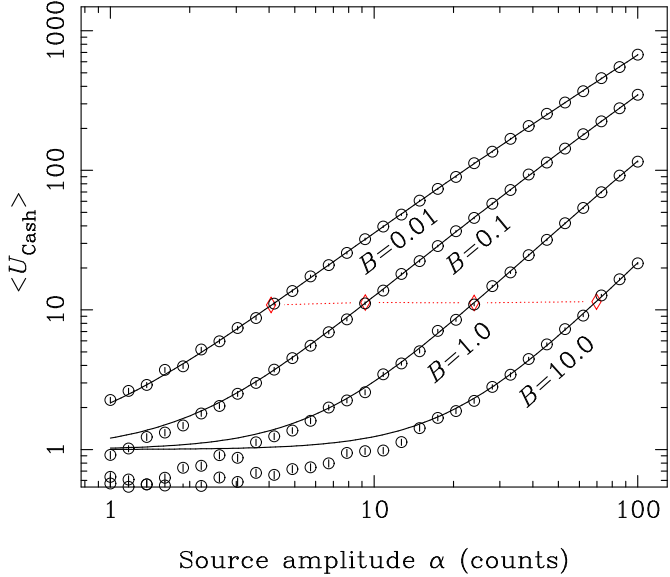}}
      \caption{Comparison between the approximate expression for $\langle U_\mathrm{Cash} \rangle$ developed in section \ref{sd4_cash_approx} and Monte Carlo estimations of it. The Monte Carlo points are indicated by circles centred on error bars (many of the error bars are too small to be visible on this plot). All values were calculated for the single band case with $M=4$. Half-tone diamonds, connected by dotted lines (both coloured red in the electronic version), indicate the sensitivities appropriate to each background value.
              }
         \label{fig_c}
   \end{figure}

As described in appendices \ref{app_PU_1band_cash}, \ref{app_PU_nband_cash} and \ref{app_PU_nband_cash_n}, no closed-form expression for $\langle U_\mathrm{Cash} \rangle$ is known. In its absence, inversion to derive $\alpha_\mathrm{det}$ from $U_\mathrm{det}$ must proceed via a tedious and problematic Monte Carlo. Such an approach is clearly going to be inadequate for `in the field' calculations of sensitivity. The following expression appears however to offer an acceptable approximation for such cases:
\begin{equation} \label{equ_f}
	\langle U_\mathrm{Cash} \rangle \sim 1 + 2 \sum_i \left( [B_i + \alpha S_i] \ln \left[\frac{B_i + \alpha S_i}{B_i} \right] \right) - 2 \alpha \sum_i S_i.
\end{equation}
This expression is compared to Monte Carlo data for several values of background in figure \ref{fig_c}.

Equation \ref{equ_f} was obtained by a series of guesses at reasonable approximations. Initially, a formula was sought for the case in which negative-amplitude fits are \emph{not} discarded. The additional `fudge factor' of unity produced a function which gave an excellent fit to the Monte Carlo data for such a case. Discarding negative fits, as is done for all results in the present paper, degrades this close match at low amplitudes. However, the approximation is seen to be a good match at $\alpha$ values corresponding to realistic sensitivities (the half-tone diamonds in figure \ref{fig_c}). For this reason it is considered to be useful for the construction of sensitivitity maps for Cash source detection.

\subsection{A Cash statistic for sliding-box} \label{sd4_cashbox}

There are two problems in the calculation of $\langle U \rangle$ and $P$ for the \emph{eboxdetect} multi-band algorithm: firstly, there appears to be no closed-form expression for $\langle U \rangle$; secondly, the best available approximation for $P$ is significantly inaccurate at low values of background.

\subsubsection{Incorrect $\langle U \rangle$} \label{sd4_cashbox_U}

In paper I, sensitivities for the \emph{eboxdetect} detection method were calculated using the assumption that $\langle U \rangle$ could be approximated by
\begin{equation} \label{equ_e}
	\langle U \rangle \sim -2 \sum_{k=1}^N \ln(P[>\!\!\langle U_k \rangle;\alpha\!=\!0]).
\end{equation}
$P(>\!\!\langle U_k \rangle;\alpha\!=\!0)$ here is given by
\begin{displaymath}
	P(>\!\!\langle U_k \rangle;\alpha\!=\!0) = 1 - Q \left(\sum_{i,j} B_{i,j,k} + \alpha \sum_{i,j} S_{i,j,k}, \sum_{i,j} B_{i,j,k} \right)
\end{displaymath}
where $Q$ is the complementary incomplete gamma function\footnote{The somewhat awkward notation $1-Q$ is here chosen, instead of the more compact and natural symbol $P = 1-Q$, representing the other incomplete gamma function, merely in order to avoid using $P$ for more than one quantity. Other alternative representations for the incomplete gamma functions alas offer no improvement since they involve additional mentions of the gamma function $\Gamma(a)$.} defined by
\begin{displaymath}
  Q(a,x) = \frac{1}{\Gamma(a)} \int_x^{\infty} dt \, \exp(-t) t^{a-1}.
\end{displaymath}
This expression was numerically inverted by use of Ridders' method to derive the sensitivity $\alpha_\mathrm{det}$ from $U_\mathrm{det}$. Monte Carlo experiments show, however, that equation \ref{equ_e} overestimates $\alpha_\mathrm{det}$ by about 10\% over most of the background range. It is simply incorrect and should not be used.

\subsubsection{Inaccurate $P$} \label{sd4_cashbox_P}

Figure 3 of paper I shows that use of the \emph{eboxdetect} formula for $P$ (ie, the formula used to calculate the values written by \emph{eboxdetect} to the LIKE column of its output source list) can be incorrect by as much as an order of magnitude at moderately low values of background. Without an understanding of the theoretical underpinnings of the \emph{eboxdetect} method, it is difficult to suggest a rigorously better alternative. All sensitivity values calculated for this method in the present paper therefore made use of a Monte Carlo approximation.

\subsubsection{A more accurate alternative} \label{sd4_cashbox_cashbox}

One way around these difficulties is to recast the \emph{eboxdetect} algorithm in terms of the Cash statistic. This probably doesn't make the method any more sensitive, but it does make its sensititivity (and its detection likelihoods) easier to calculate.

Since we are not using the information in the spatial variation of $S$, both background and counts can be added across the detection field to give, respectively, a single number for each band. We define $U_\mathrm{BoxCash}$ as
\begin{displaymath}
	U_\mathrm{BoxCash} = \sum_{k=1}^N U_k
\end{displaymath}
where $U_k$ has the usual Cash form, even though there is here only one `pixel' to sum over:
\begin{displaymath}
	U_k = 2 \ c_{\mathrm{total},k} \ \ln \left(\frac{B_{\mathrm{total},k} + \hat{\alpha}_k S_{\mathrm{total},k}}{B_{\mathrm{total},k}} \right) - 2 \hat{\alpha}_k S_{\mathrm{total},k}
\end{displaymath}
for
\begin{displaymath}
	B_{\mathrm{total},k} = \sum_{i=-M}^{M} \sum_{j=-M}^{M} B_{i,j,k}
\end{displaymath}
etc. In this instance it is easy to deduce that, for the $k$th band,
\begin{displaymath}
	\hat{\alpha}_k = \frac{c_{\mathrm{total},k} - B_{\mathrm{total},k}}{S_{\mathrm{total},k}};
\end{displaymath}
The expression for $U_k$ thus becomes
{\setlength\arraycolsep{2pt}
\begin{eqnarray*}
	U_k & = & B_{\mathrm{total},k} - C_{\mathrm{total},k} \ln \left(\frac{C_{\mathrm{total},k}}{B_{\mathrm{total},k}} \right) \mathrm{\ where \ } C_{\mathrm{total},k} > 0 \\
	    & = & B_{\mathrm{total},k} \mathrm{\ otherwise.}
\end{eqnarray*}}
$P$ is as given in the appendix \ref{app_PU_nband_cash_n}; and we can now use the approximation in equation \ref{equ_f} to estimate (and therefore invert) $\langle U \rangle$. Monte Carlos show that the Cash $\chi^2$ prediction remains a good fit to $P$, with the usual $\times 0.5$ normalization.

\section{Conclusions}

This paper has attempted to set out clearly the issues involved in the detection of sources in images which exhibit Poisson noise. A three-stage detection algorithm was proposed in which the first stage makes a map of the detection likelihood at each image pixel, the second searches for peaks in this map and the third fits a source profile to each peak so as to return best-fit values and uncertainties for the amplitude and position of each source. Some conditions which the data must obey in order for this approach to succeed were examined. The proper way to define detection sensitivity was also rather carefully gone into.

It was realized after the issue of paper I that the probability distribution of map values is not in general the same as the distribution of values at a set of pixels restricted to peaks in that map. This has implications for the calculation of the sensitivity of a detection method, since the starting point in this calculation, namely the acceptable maximum in the number density of false detections returned by the method, translates to a cutoff value in the peak rather than the map distribution. On the other hand, because the map and peak distributions can be shown to converge at high values, it was felt allowable to use the map distribution as a proxy for the peak one when comparing sensitivities of different detection methods. Thus the results of paper I are not invalidated and could be extended here.

The Cash likelihood-ratio statistic (LRS) as a means of detecting sources amid Poissonian noise was here described. This statistic has been used for source detection at least in Rosat images (Cruddace et al \cite{cruddace_1977}, and Boese and Doebereiner \cite{boese_2001}) and also for XMM-Newton (Watson et al \cite{1xmm_paper}, \cite{2xmm_paper}). According to Cash (\cite{cash_1979}), or rather to Wilks (\cite{wilks_1963}) as cited by Cash, such an LRS ought to be distributed as $\chi^2$ with $q$ degrees of freedom, where $q$ is the number of free parameters. It has however been claimed that the LRS is not well suited to source detection because the probability distribution of values of such a statistic calculated under the null hypothesis (ie that the image contains no sources) is not well known. Protassov et al (\cite{protassov_2002}) for example maintain that the source-detection application of LRS disobeys some of the necessary regularity conditions under which the distribution of an LRS can be expected to follow that of $\chi^2$. In the present paper this claim is disputed; Monte Carlo experiments are also described which show, under typical x-ray instrument conditions, and provided that sources with negative fitted amplitude are discarded from the ensemble, that the distribution of the Cash statistic, evaluated for images simulated from a source-free flux model, adheres in shape at least quite closely to a $\chi^2$ distribution with the appropriate degrees of freedom. All that is necessary is to evaluate the normalization constant between the two.

Although it was found convenient for purposes of the present paper to write bespoke code to perform the Cash-statistic source detection and fitting, the XMM-Newton SAS task \emph{emldetect} was briefly compared to this bespoke code, and its results shown to also obey the $\chi^2$ distribution.

The main thrust of the present paper has been to compare the sensitivity of different detection methods. The Cash-statistic detection is shown to be the best of the methods tested. In terms purely of sensitivity it is seen to be only marginally superior to an optimized linear filter; however the Cash method is shown to possess two further advantages. Firstly, it is more flexible, since it allows the user the option of assuming a source spectrum or leaving this free. Secondly, for the final detection probability, one need not just accept the `map' value at the pixel nearest to the final source position: instead one can calculate a final, accurate value specific to that fitted position.

The flexibility of the Cash statistic was also demonstrated by showing how it can be adapted to calculate detection likelihoods for sliding-box source detection as performed for example by the SAS task \emph{eboxdetect}. The advantage in doing so is that one can make use of the demonstrated close agreement between the distribution of the Cash statistic and $\chi^2$. This would represent a considerable improvement over the likelihood formula employed within \emph{eboxdetect}.

As a final result, the paper describes approximations which can be used to calculate sensitivity maps appropriate to Cash-style source detection. For XMM-Newton the only previous ways to do this have been via the SAS task \emph{esensmap} (which examination of the code shows contains an algorithm which could only have serendipitous success), or via an empirical correction to sliding-box sensitivities (eg Carrera et al \cite{carrera_2007}).

\begin{acknowledgements}
Thanks are due firstly to Georg Lamer for providing the vital clue to understanding the Cash algorithm; and secondly to Anja Schr\"oder for valuable assistance in obtaining XMM-Newton data.
\end{acknowledgements}

\appendix

\section{Cash amplitude fitting} \label{app_cash}

To calculate a map of the Cash statistic (section \ref{sd2_cash_def}) in a single spectral band, we centre a PSF $S$ on each pixel of the image and fit only its amplitude $\alpha$. The fit is performed over a small number $N$ pixels surrounding the pixel for which we are calculating the value. $L_\mathrm{null}$ is calculated by setting $\alpha$ to zero. The resulting probability is then the probability of obtaining the fitted value $\hat{\alpha}$ when there is in fact nothing there but background $B$. The expression, derived from equation \ref{equ_a}, for the Cash statistic relevant to this case is
\begin{equation} \label{equ_d}
	U_\mathrm{Cash} = 2 \sum_{i=1}^N c_i \ln \left(\frac{B_i + \hat{\alpha} S_i}{B_i} \right) - 2 \hat{\alpha} \sum S_i
\end{equation}
where $\hat{\alpha}$ is the fitted value of $\alpha$.

As mentioned in section \ref{sd2_cash_objections}, statistically speaking it is permitted for $\hat{\alpha}$ to be negative; whether we later discard such occurrences depends on the physical model which is appropriate, and also on what we wish to do with the data. What is \emph{not} permitted is for $B_i + \hat{\alpha} S_i$ to be negative for any $i$. (If the recorded counts at the $i$th pixel is $>0$, $B_i + \hat{\alpha} S_i$ is not permitted to equal zero either - but this is a formal condition which has no effect on the computational practicalities.) The absolute limit on permitted $\hat{\alpha}$ values in the negative direction is thus $-\min(B/S)$.

\begin{figure}
   \centering
      \resizebox{\hsize}{!}{\includegraphics[angle=-90]{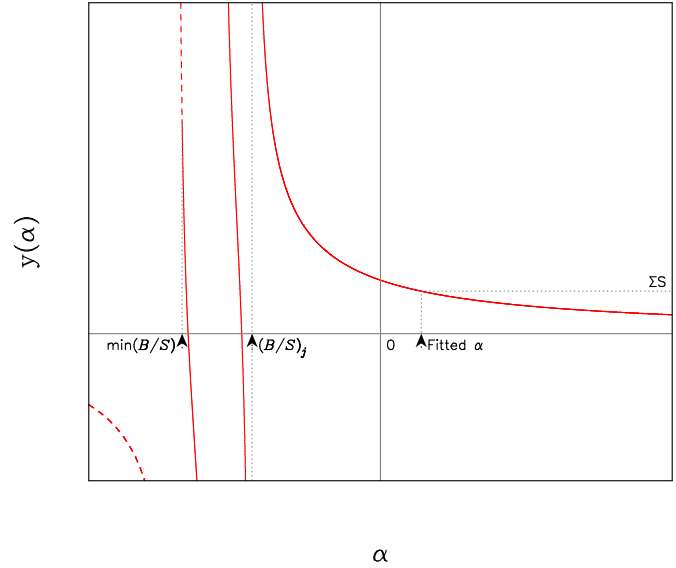}}
      \caption{An example plot of equation \ref{y_func} with 3 singularities. The dashed portion represents values beyond the `statistical limit'.
              }
         \label{fig_r}
   \end{figure}

If $c_i = 0$ for all $N$ pixels in the fitted patch, $\hat{\alpha}$ should be set to this limit. This, the model with the least amount of flux which the parameter limits permit, is the one most likely to explain the lack of counts. If some $c_i > 0$ though, of course $\hat{\alpha}$ may be larger than $-\min(B/S)$. This possibility can be explored by taking the derivative of $U$ with respect to $\hat{\alpha}$ and setting it to zero. This gives
\begin{equation} \label{equ_b}
	\frac{\partial U_\mathrm{Cash}}{\partial \hat{\alpha}} = 2 \sum_{i=1}^N \left(\frac{c_i}{\hat{\alpha} + B_i / S_i} \right) - 2 \sum S_i = 0.
\end{equation}
Equation \ref{equ_b} can be solved numerically for $\hat{\alpha}$, using for example the fast and robust Newton-Rhapson method (cf Press et al \cite{numerical_recipes}). To help in visualizing how to apply this it is useful to look at a diagram. Figure \ref{fig_r} shows an example plot of the function
\begin{displaymath} \label{y_func}
	y(\hat{\alpha}) = \sum_{i=1}^N \left(\frac{c_i}{\hat{\alpha} + B_i / S_i} \right).
\end{displaymath}
$y$ has $M$ hyperbolic singularities, one for each of the subset of $M$ values of $c_i$ which are $> 0$. Although formally speaking there are $M$ solutions to equation \ref{equ_b} (indicated in figure \ref{fig_r} by intersections of the graph with the horizontal line at $y = \sum S_i$), the largest solution is clearly the one desired.

Newton-Rhapson solution can be facilitated in two ways: by estimating bounds for $\hat{\alpha}$ and by performing a coordinate transform. Firstly, let us label as the $j$th that pixel for which the following conditions hold: firstly that $c > 0$ for that pixel and secondly that, of that subset of $M$ pixels for which this holds, $B/S$ is the smallest. Clearly the most positive singularity occurs at $\hat{\alpha} = -(B/S)_j$. $\hat{\alpha}$ is bounded as follows:
\begin{displaymath}
	\frac{c_j}{\sum S_i} \le \hat{\alpha} + B_j/S_j \le \frac{\sum c_i}{\sum S_i}.
\end{displaymath}

The rate of convergence of the Newton-Rhapson algorithm can be accelerated by performing a coordinate transform
\begin{displaymath}
	u = 1/(\hat{\alpha} + B_j/S_j).
\end{displaymath}

The `statistical' bounding condition on $\hat{\alpha}$, namely $\hat{\alpha} > -\min(B/S)$, must be applied if the Newton-Rhapson solution exceeds it, as is possible.

\section{Formulae for sensitivity calculations} \label{app_PU}
\subsection{1-Band detection} \label{app_PU_1band}

Expressions for $U$, $P(>\!\!U;0)$ and $\langle U \rangle$ are given below for the three single-band detection methods compared in section \ref{sd3_1band_comp}. The $k$ index is superfluous in the present single-band scenario and has therefore been suppressed for the time being.

\subsubsection{1-band sliding-box} \label{app_PU_1band_box}

\begin{equation} \label{equ_1band_box_u}
	U = \sum_{i=-M}^{M} \sum_{j=-M}^{M} c_{i,j}.
\end{equation}
\begin{displaymath}
	P(>\!\!U;0) = 1 - Q(U, [2M+1]^2 \times B_{i,j}),
\end{displaymath}
where $Q$ is the complementary incomplete gamma function as previously.
\begin{displaymath}
	\langle U \rangle = \sum_{i=-M}^{M} \sum_{j=-M}^{M} B_{i,j} + \alpha \sum_{i=-M}^{M} \sum_{j=-M}^{M} S_{i,j}.
\end{displaymath}

\subsubsection{1-band optimized linear filter} \label{app_PU_1band_matched}

\begin{displaymath}
	U = \sum_{i=-M}^{M} \sum_{j=-M}^{M} w_{i,j} c_{i,j},
\end{displaymath}
where the $w_{i,j}$ are optimum weights, calculated by a procedure described in paper I.
\begin{displaymath}
	P(>\!\!U;0) \sim Q \left( \frac{B^\prime}{w_\mathrm{equiv}}, \frac{U}{w_\mathrm{equiv}} \right)
\end{displaymath}
where
\begin{displaymath}
  B^\prime = \sum_{i=-M}^{M} \sum_{j=-M}^{M} w_{i,j} B_{i,j}
\end{displaymath}
and
\begin{displaymath}
  w_\mathrm{equiv} = \frac{\sum_{i=-M}^{M} \sum_{j=-M}^{M} w_{i,j}^2 B_{i,j}}{\sum_{i=-M}^{M} \sum_{j=-M}^{M} w_{i,j} B_{i,j}}.
\end{displaymath}
\begin{displaymath}
	\langle U \rangle = \sum_{i=-M}^{M} \sum_{j=-M}^{M} w_{i,j} (B_{i,j} + \alpha  S_{i,j}).
\end{displaymath}

\subsubsection{1-band Cash-statistic detection} \label{app_PU_1band_cash}

Expansion of equation \ref{equ_d} to include the extra spatial index gives
\begin{equation} \label{equ_i}
	U = 2 \sum_{i=-M}^{M} \sum_{j=-M}^{M} c_{i,j} \ln \left(\frac{B_{i,j} + \hat{\alpha} S_{i,j}}{B_{i,j}} \right) - 2 \hat{\alpha} \sum_{i=-M}^{M} \sum_{j=-M}^{M} S_{i,j}.
\end{equation}
\begin{displaymath}
	P(>\!\!U;0) \sim Q(1/2, U/2).
\end{displaymath}
There appears to be no easy route to an exact, closed-form expression giving $\langle U_\mathrm{Cash} \rangle$ in terms of $B$, $\alpha$ and $S$ (although see section \ref{sd4_cash_approx} for an approximate formula). One can estimate the $\alpha$ appropriate to a given value of $\langle U \rangle$ via iterative Monte Carlos as described in section \ref{sd3_technique}; in fact this must be done if one wants to investigate the effect on the Cash detection algorithm of fitting with the wrong shape $S$. All the sensitivity values for the Cash algorithm presented in the present paper have therefore been estimated by this method.

\subsection{$N$-band detection} \label{app_PU_nband}

Expressions for $U$, $P(>\!\!U;0)$ and $\langle U \rangle$ are given below for the five single-band detection methods compared in section \ref{sd3_nband_comp}.

\subsubsection{Sliding-box detection on a single, all-band image} \label{app_PU_nband_box}

This is exactly the same as the algorithm described in section \ref{app_PU_1band_box}. The sum over $N$ bands is here just made explicit.

This method does not require the user to choose a source spectrum template.
\begin{displaymath}
	U = \sum_{i=-M}^{M} \sum_{j=-M}^{M} \sum_{k=1}^N c_{i,j,k},
\end{displaymath}
\begin{displaymath}
	P(>\!\!U;0) = 1 - Q(U, [2M+1]^2 \times \sum_{k=1}^N B_{i,j,k}),
\end{displaymath}
and
\begin{displaymath}
	\langle U \rangle = \sum_{i=-M}^{M} \sum_{j=-M}^{M} \sum_{k=1}^N B_{i,j,k} + \alpha \sum_{i=-M}^{M} \sum_{j=-M}^{M} \sum_{k=1}^N S_{i,j,k}.
\end{displaymath}

\subsubsection{Sliding-box detection, \emph{eboxdetect} statistic} \label{app_PU_nband_box_n}

This is the detection method employed by the XMM-Newton SAS task \emph{eboxdetect} (online description at http://xmm.vilspa.esa.es/sas/current/doc/eboxdetect.ps.gz). The method does not require the user to choose a source spectrum template.

$U$ is calculated in a three-step process. In the first step, counts within the box are summed for each $k$th band to give $U_k$. The second step forms $P(>\!\!U_k;0)$ for each $k$. The final $U$ is then given by
\begin{equation} \label{equ_eboxdetect_U}
	U = \sum_{k=1}^N L_k
\end{equation}
where
\begin{displaymath}
	L_k = -2 \ln(P[>\!\!U_k;\alpha\!=\!0]).
\end{displaymath}
No description of \emph{eboxdetect} appears to have been published; hence the theoretical basis of this expression is obscure.

If one assumes (which is not strictly true) that the $L_k$ above follow distributions which don't change with $k$, one can apply the Central Limit Theorem to equation \ref{equ_eboxdetect_U} to give, in the limit of high $N$, the following form\footnote{This derivation makes use of the identity $\mathrm{erfc}(x) = Q(0.5,x^2)$.} for $P$:
\begin{displaymath}
	P(>\!\!U;0) = 0.5 (1 + Q[0.5, U^2/2 N \sigma^2_L])
\end{displaymath}
where $\sigma^2_L$ is the variance of $L_k$ for any $k$. However in \emph{eboxdetect} practice, still evident in the code in \emph{eboxdetect} version 4.19, the actual expression used is
\begin{equation} \label{equ_h}
	P(>\!\!U;0) \sim Q(N, U/2);
\end{equation}
Some experimentation in both paper I and the present paper confirms that equation \ref{equ_h} does appear to match Monte Carlo results in the limit of high background. The issue is in any case academic for present purposes, since as in all cases where only an approximate expression for $P$ is available, the results presented here are derived by use of a Monte Carlo estimation of $P$.

No closed-form expression for $\langle U \rangle$ exists: it must be calculated numerically.

\subsubsection{$N$-band optimized linear filter} \label{app_PU_nband_matched}

Essentially this is the same algorithm as described in section \ref{app_PU_1band_matched}, just with the sum over the $N$ bands made explicit. This method requires the user to choose a source spectrum template, in order to calculate the optimum weights $w$.
\begin{displaymath}
	U = \sum_{i=-M}^{M} \sum_{j=-M}^{M} \sum_{k=1}^N w_{i,j,k} c_{i,j,k};
\end{displaymath}
$P$ is the same, but with of course
\begin{displaymath}
  B^\prime = \sum_{i=-M}^{M} \sum_{j=-M}^{M} \sum_{k=1}^N w_{i,j,k} B_{i,j,k}
\end{displaymath}
and
\begin{displaymath}
  w_\mathrm{equiv} = \frac{\sum_{i=-M}^{M} \sum_{j=-M}^{M} \sum_{k=1}^N w_{i,j,k}^2 B_{i,j,k}}{\sum_{i=-M}^{M} \sum_{j=-M}^{M} \sum_{k=1}^N w_{i,j,k} B_{i,j,k}};
\end{displaymath}
\begin{displaymath}
	\langle U \rangle = \sum_{i=-M}^{M} \sum_{j=-M}^{M} \sum_{k=1}^N w_{i,j,k} (B_{i,j,k} + \alpha  S_{i,j,k}).
\end{displaymath}

\subsubsection{Cash-statistic detection with a spectrum template} \label{app_PU_nband_cash}

Essentially this is the same algorithm as described in section \ref{app_PU_1band_cash}, just with the sum over the $N$ bands made explicit. This algorithm requires the user to choose a source spectrum template.
\begin{displaymath}
	U = 2 \! \sum_{i=-M}^{M} \sum_{j=-M}^{M} \sum_{k=1}^N c_{i,j,k} \ln \left(\frac{B_{i,j,k} + \hat{\alpha} S_{i,j,k}}{B_{i,j}} \right) - 2 \hat{\alpha} \! \sum_{i=-M}^{M} \sum_{j=-M}^{M} S_{i,j}.
\end{displaymath}
$P$ is the same as in section \ref{app_PU_1band_cash}, and $\langle U \rangle$ is again estimated via a Monte Carlo procedure (though see section \ref{sd4_cash_approx}).

\subsubsection{Cash-statistic detection without a spectrum template} \label{app_PU_nband_cash_n}

The difference between this method and the preceding one is that here a value of $\hat{\alpha}$ is calculated, ie the source PSF is fitted to the data, separately for each band. No knowledge of the source spectrum is required. $U$ is then formed by summing the $U_k$ for each band. Since there are now $N$ degrees of freedom in the fit, $P$ is here given by
\begin{displaymath}
	P(>\!\!U;0) \sim Q(N/2, U/2).
\end{displaymath}
$\langle U \rangle$ is again estimated via a Monte Carlo procedure (though see section \ref{sd4_cash_approx}).

\end{document}